# A four-pixel single-photon pulse-position camera fabricated from WSi superconducting nanowire single-photon detectors

V. B. Verma[1*], R. Horansky[1], F. Marsili[2], J. A. Stern[2], M. D. Shaw[2], A. E. Lita[1], R. P. Mirin[1], and S. W. Nam[1]

[1]*National Institute of Standards and Technology, 325 Broadway, Boulder, CO 80305, USA*

[2]*Jet Propulsion Laboratory, 4800 Oak Grove Dr., Pasadena, California 91109, USA*

We demonstrate a scalable readout scheme for an infrared single-photon pulse-position camera consisting of WSi superconducting nanowire single-photon detectors (SNSPDs). For an $N \times N$ array, only $2 \times N$ wires are required to obtain the position of a detection event. As a proof-of-principle, we show results from a $2 \times 2$ array.

Arrays of single-photon detectors have potential applications in infrared spectroscopy[1] and optical communications in the so-called "photon starved" regime.[2,3] One example is interplanetary communications where the mass and power limitations of the laser transmitter result in single-photon light levels at the receiver. In such a regime, the photon information efficiency measured in bits/photon becomes an important metric determining the capacity of the communications channel. One technique to convey more than one bit per photon is pulse-position modulation (PPM), in which $M$ bits are encoded by a single photon transmitted in one of $2^M$ possible time-shifts or spatial modes.[3-5] By using a transmitter in which the light source is spatially modulated, an $N \times N$ single photon detector array in which the position of the single photon can be extracted could be used to transmit more than one bit per photon. Superconducting nanowire single-photon detectors (SNSPDs) are ideal for this application due to their fast recovery times ($< 100$ ns),[6,7] low jitter ($< 100$ ps),[8] low dark count rates ($< 1$ count per second, cps),[9] and high efficiencies ($> 90\%$).[9]

We present a scalable readout scheme for an array of SNSPDs that yields position-dependent information about where in the array the photon was absorbed. Although there have been other reports of incorporating SNSPDs into arrays,[10-12] these approaches are targeted at achieving high count rates, and do not yield position-dependent information about where the photon detection event occurred. Due to the fact that SNSPDs are operated at cryogenic temperatures, the number of coaxial cables used to read-out and bias the detectors is restricted by the cooling power of the cryogenic system. Thus, a scheme in which each detector is read-out individually quickly becomes impractical as the number of detectors is scaled. The scheme we present requires only $2N$ coaxial cables for an $N \times N$ array, greatly reducing the cryogenic heat load requirements of the system while maintaining the high timing accuracy of the detectors. As a proof-of-principle demonstration, we show results from a 2 × 2 array.

To demonstrate the scalability of the proposed scheme, Fig. 1(a) shows a circuit diagram for a 3 × 3 array. The SNSPDs are biased with a current $I_B$ through a bias-T in each column. Each pixel consists of a series resistor $R_S$ followed by the SNSPD (with kinetic inductance $L_k$) and an additional inductor $L_i$. In the diagram the SNSPD and additional inductor are lumped into one element represented by a circle. In simplified terms, each SNSPD can be represented as an electrical switch in parallel with a resistor $R_N$. When the SNSPD is in the superconducting state, the electrical switch is closed, shorting the resistor $R_N$. In this steady state situation, the circles in Fig. 1(a) are equivalent to short circuits, and the bias current $I_B$ is equally distributed between all three pixels in the column. When a photon is absorbed by one pixel, a normal region or "hotspot" is generated in the nanowire, causing the electrical switch to open, and diverting the current into $R_N$.[13] In this case the circles in Fig. 1(a) can be replaced with resistors of magnitude $R_N \gg R_S = R_R = 25\ \Omega$ in series with the total pixel inductance $L_k+L_i$. Typically the value of $R_N$ is on the order of 1 k$\Omega$ for SNSPDs fabricated from amorphous WSi. The high pixel resistance causes the current through that pixel to be diverted to the bias-T and amplifiers where a voltage pulse is produced in the column readout. The current through the row resistor $R_R$ is simultaneously reduced, resulting in a negative voltage pulse in the row readout.

We note that when the SNSPD becomes resistive, some of the current may also be diverted to adjacent pixels in the column. Current leakage adjacent pixels is undesirable since it can either cause them to switch to the normal state if they are biased close to their switching currents, or it can cause fluctuations in the detection efficiency (assuming the detection efficiency is a function of bias current up to the switching current, which is common in NbN-based SNSPDs but not WSi SNSPDs as will be discussed later). Current leakage also reduces the signal-to-noise ratio and increases the jitter of the voltage pulse produced by the amplifier for the pixel that detects the photon. The purpose of the additional kinetic inductance $L_i$ in series with each SNSPD is to minimize this leakage current. Finally, as the resistive SNSPD pixel cools to the superconducting state, the series resistance $R_S$ ensures that current is equally redistributed amongst all pixels in the column.

The pulse height, signal-to-noise ratio, and hence the jitter of the row pulse scales in proportion to $R_R$. Thus it is desirable to make $R_R$ as large as possible without causing the SNSPDs to latch into the normal state due to speedup of the current recovery into the pixel.[14, 15] For the 4-pixel array presented here, we found that 25 Ω was large enough to accurately discriminate the row pulse without causing the SNSPDs to latch. For the series resistor $R_S$ there is no gain in performance when the value of $R_S$ is increased. Since each pixel will experience a slightly different total resistance due to variations in the lengths of the gold wiring leads, the value of $R_S$ need only be larger than the resistance of the wiring leads across the entire span of the array so that each pixel experiences a similar resistance. This is important to ensure that each pixel recovers to the same value of current bias after a photon detection event.

In this work we demonstrate a 2 × 2 array as a proof-of-principle of the array readout architecture. Fig. 1 (b) shows a detailed electrical diagram of the circuit, which is identical in concept to Fig. 1 (a). Each circle used to represent a pixel in Fig. 1(a) has been expanded into its electrical equivalent of a total inductance ($L_{tot} = L_k + L_i$) plus the parallel combination of a switch and a normal state resistance $R_N$. All components in the diagram are patterned on-chip, except for the bias-T's and amplifiers, which are located outside of the cryostat at room temperature. The total gain of the amplifier chain is 51.5 dB. For simplicity and to aid with the following discussion, we refer to the top row, bottom row, right column, and left column using the cardinal directions (North, South, East, and West, respectively).

The SNSPD array is fabricated on a silicon wafer with 150 nm of thermally-grown $SiO_2$ on top. Contact pads and resistors are patterned and deposited in the same step, and consist of 2 nm Ti and 50 nm Au. The wafer is then cleaned in an $O_2$ plasma and the superconducting WSi film (~ 25% Si composition, 4.6 nm thick, $T_c$ ~3.4 K) is deposited by DC magnetron cosputtering from separate W and Si targets at room temperature. Amorphous WSi has a number of properties that make it a desirable superconducting material for fabrication of a large array of SNSPDs. [9, 16, 17] WSi SNSPDs have demonstrated record-high efficiencies (93 % at 1550 nm). In addition, the reduced carrier density and larger hotspot size in WSi allows the nanowires to be wider than NbN-based nanowires, which considerably improves device yield due to a lower probability of constriction. WSi SNSPDs also show a saturation of the internal detection efficiency over a wide bias range, which is ideal for arrays since it allows each device to be current-biased well below the switching current without sacrificing efficiency. As discussed above, biasing close to the switching current can lead to undesirable firing of adjacent pixels in a column due to current leakage.

After deposition of the WSi, the layer is patterned by use of optical lithography into a 20 μm – wide stripe on which the SNSPD will be patterned. The additional inductor (3.5 times the inductance of the SNSPD) is also patterned in this step, which consists of 2 μm – wide wires on a 6 μm pitch. Following optical lithography the WSi film is etched in an $SF_6$ plasma. Finally, electron-beam lithography and $SF_6$ etching are used to pattern the SNSPDs into the 20 μm – wide strips of WSi. The SNSPDs are composed of meandering 160 nm – wide nanowires with a pitch of 360 nm. The total active area of each device is 16 μm × 16 μm. Fig. 2 shows an optical microscope image of the array showing the large inductor ($L_i$), series resistor ($R_S$), and SNSPDs. The SNSPDs are current-biased through the bias lines for each column on the left and right sides of the image.

The array was cooled to a temperature of 250 mK in an adiabatic demagnetization refrigerator (ADR) for measurement of the switching current and optical response. The device was flood-illuminated at a wavelength of 1550 nm by a single-mode optical fiber positioned ~ 8 mm away from the chip. Fig. 3 shows the total photo-count rate (PCR) and dark count rate (DCR) for the West and East columns. For this measurement, the counting electronics were triggered on the West (or East) column amplifier output. Thus, the count rate is the sum of the count rates of both SNSPDs in the same column.

The West column has a switching current ($I_{SW}$) of 15 µA, with a cutoff current ($I_{co}$, the current at the inflection point of the PCR vs. bias curve) of 9 µA. The DCR is < 1 cps except above 90 % of $I_{SW}$, where it slowly increases to ~ 1 kcps at $I_{SW}$. The East column has a slightly higher switching current than the West column of 18.9 µA, which could be attributed to a constriction in one of the SNSPDs or the large inductor in the West column. The DCR for the East column is < 1cps below 80 % of $I_{SW}$, and increases to 25 cps at $I_{SW}$. Also note that the maximum count rate for the West column is approximately twice the maximum count rate of the East column, which is likely due to the chip being misaligned relative to the fiber so that the West column experiences a higher photon flux. The higher PCR for the West column is also consistent with the higher DCR, which supports the claim that the DCR is due primarily to blackbody photons coupled into the fiber and not intrinsic dark counts from the detector. [9, 16, 17]

Fig. 4 shows averaged voltage pulse traces for the North, South, East, and West amplifier outputs triggered on single-photon detection events in the (a) NE, (b) NW, (c) SE, and (d) SW quadrants of the array. To obtain these traces the oscilloscope is set to trigger on the logical AND (^) of two of the four inputs (i.e., (a) N ^ E, (b) N ^ W, etc.). Bias currents of the West and East columns were set at 13 µA and 18 µA, respectively. As shown in the plots, voltage pulses from the North and South row outputs are negative with a peak of 60 mV, and a $1/e$ decay time of 34 ns. Pulses from the West and East column outputs are positive with magnitudes of 180 mV and 246 mV, and $1/e$ decay times of 24 ns. Note that when a negative pulse is observed on a row output (i.e., North in Fig. 1 (a)), a very small positive pulse is observed on the other row output (South in Fig. 1 (a)). This is due to the small current leakage into the adjacent non-firing column pixel, which is minimized by the extra kinetic inductance ($L_i$) added in series with each pixel.

In summary, we have demonstrated a readout scheme for an array of SNSPDs in which the spatial position of the pixel that detects a photon is identified. For an N × N array, only 2N coaxial cables are required for read-out, which significantly reduces the cooling power requirements in a cryogenic system. As a proof-of-principle, we fabricated and tested a 2 × 2 array, although the scheme can be readily scaled to larger arrays.

This work was supported by the DARPA InPho program.

**Figures**

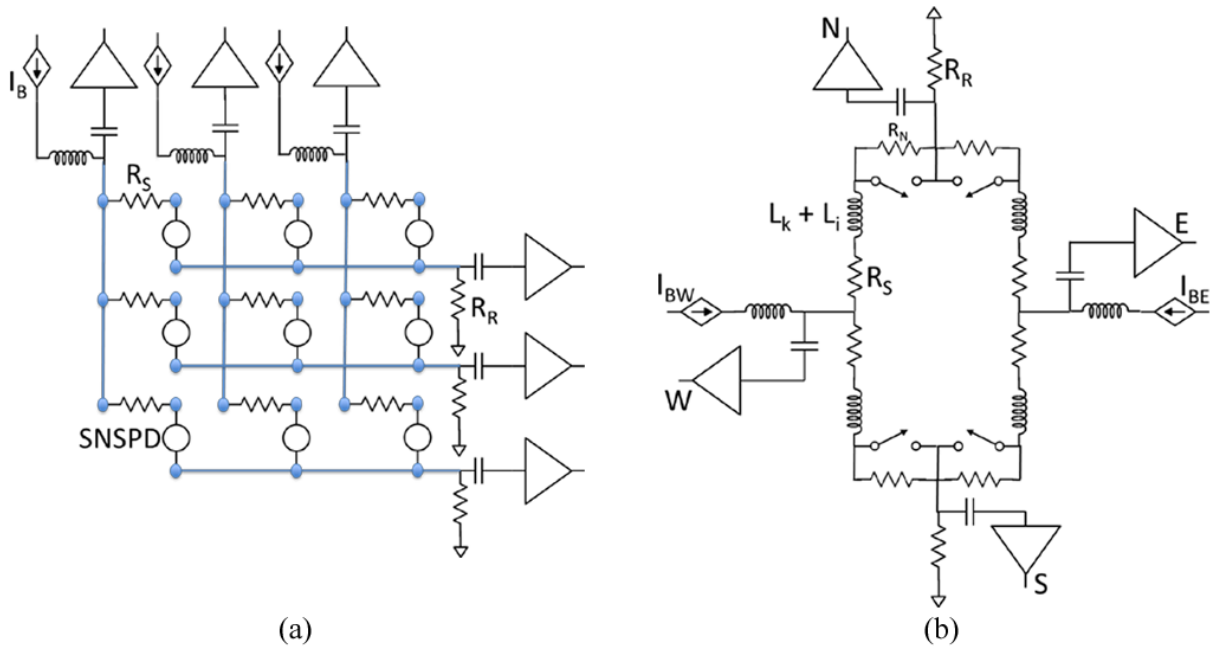

(a)                 (b)

Fig. 1 (a) Circuit diagram illustrating the scalability of the detection scheme with a 3 × 3 array of SNSPDs and (b) detailed circuit diagram for a 2 × 2 SNSPD array.

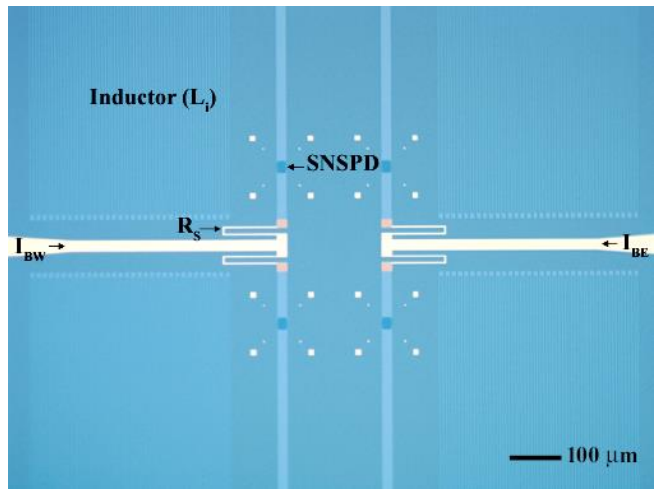

Fig. 2 Optical microscope image of the 2 × 2 SNSPD array.

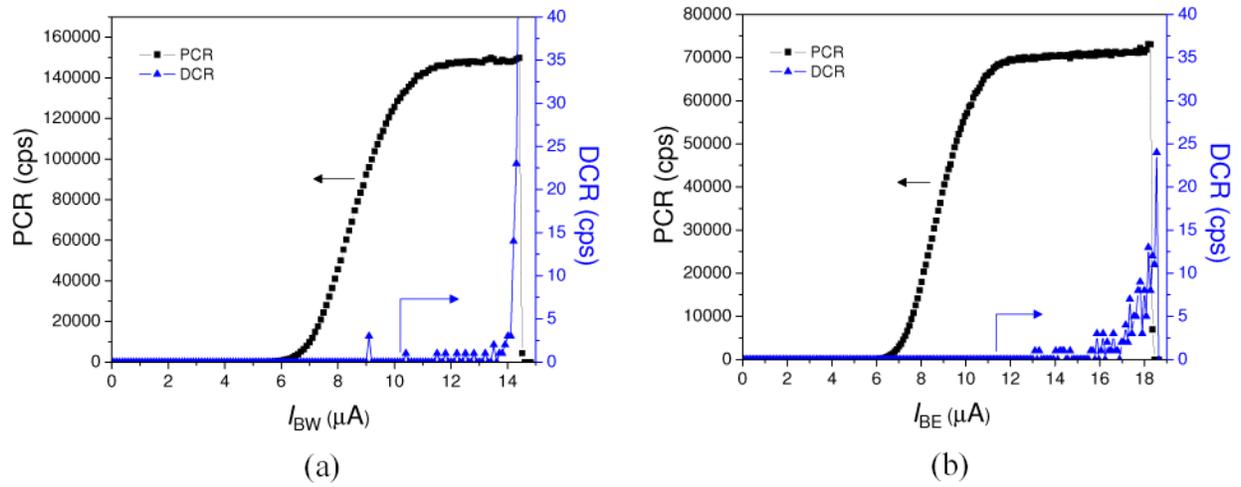

Fig. 3 Photo count rate (PCR) and dark count rate (DCR) as a function of bias current ($I_B$) for the (a) West and (b) East column bias lines.

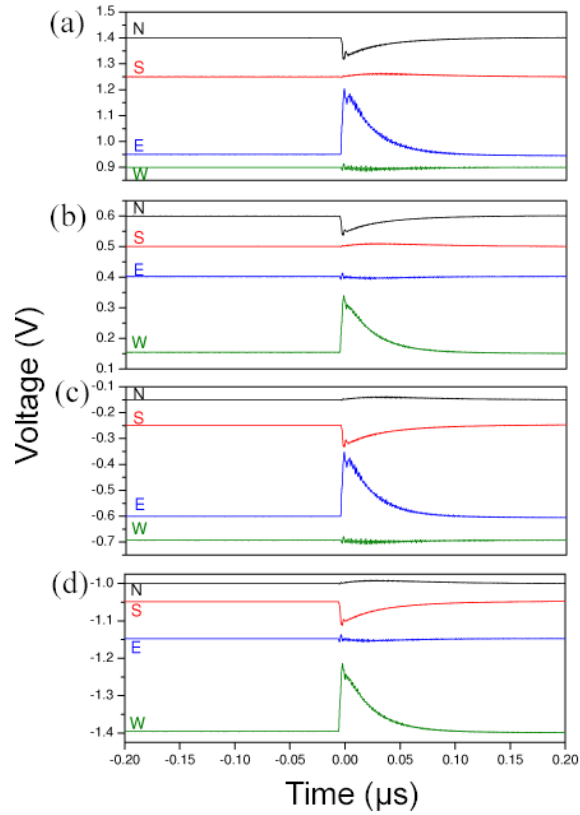

Fig. 4 Average traces of the voltage pulses from the N, S, E, and W outputs of the array triggered on the logical AND (^) of (a) N ^ E, (b) N ^ W, (c) S ^ E, (d) S ^ W. This corresponds to single photon detection events in the (a) NE, (b) NW, (c) SE, and (d) SW quadrants of the array.